\documentclass[aps,pre,amsmath,amssymb,twocolumn,showpacs,byrevtex,superscriptaddress]{revtex4}
\usepackage{graphicx}
\begin{document}
\title{Dynamics in a supercooled liquid of symmetric dumbbells:\\
Reorientational hopping for small molecular elongations}

\author{Angel J. Moreno}
\affiliation{\hspace{-1mm}Dipartimento di Fisica and INFM-CRS-SOFT, Universit\`a di Roma {\em La Sapienza}, P.le. A. Moro 2, 00185 Roma, Italy}
\affiliation{Donostia International Physics Center, Paseo Manuel de Lardizabal 4, 20018 San Sebasti\'{a}n, Spain}
\author{Song-Ho Chong}
\affiliation{Institute for Molecular Science, Okazaki 444-8585, Japan}
\author{Walter Kob}
\affiliation{Laboratoire des Collo\"{i}des, Verres et Nanomat\'{e}riaux, Universit\'{e} Montpellier II and UMR 5587 CNRS, 34095 Montpellier Cedex 5, France}
\author{Francesco Sciortino}
\affiliation{\hspace{-1mm}Dipartimento di Fisica and INFM-CRS-SOFT, Universit\`a di Roma {\em La Sapienza}, P.le. A. Moro 2, 00185 Roma, Italy}

\begin{abstract}
We present extensive molecular dynamics simulations of  a liquid of symmetric dumbbells, for constant packing fraction, 
as a function of temperature and molecular
elongation. For large elongations, translational and rotational degrees of freedom freeze at 
the same temperature. For small elongations only the even rotational degrees of freedom remain coupled to translational 
motions and arrest at a finite common temperature.  The odd rotational degrees of freedom remain ergodic at all 
investigated temperature and
the temperature dependence of the corresponding characteristic time is well described by an Arrhenius law.
Finally, we discuss the evidence in favor of the presence 
of a type-$A$ transition temperature for the odd rotational degrees of freedom, distinct from the type-$B$ transition 
associated with the arrest of the translational and even rotational ones, as predicted by the mode-coupling theory for the glass transition.

\end{abstract}

\maketitle
\vspace{-1 mm}
The ideal mode-coupling theory (MCT) equations 
have been recently solved in the site-site representation for a system
of symmetric hard dumbbell molecules \cite{chongtheor1,chongtheor2}, as a function of the packing 
fraction $\varphi$ and the elongation $\zeta$. 
Interestingly enough, the theory predicts 
two different dynamic arrest scenarios, on varing 
$\varphi$ and $\zeta$ (see Fig.~1 in Ref. \cite{chongtheor1}). For large elongations, it is predicted that all rotational
correlation functions are strongly coupled 
to the translational degrees of freedom and dynamic arrest takes place at a common 
$\varphi$ value, $\varphi_c^B(\zeta)$. According to MCT, the transition
is of type-$B$, i.e. the long time limit of translational and rotational correlation functions jumps 
discontinuosly from zero to a finite value at the  ideal glass transition line.
The ideal glass transition line (the $B$-line)  has a non-monotonic shape in the $\varphi-\zeta$ plane, 
with a maximum at $\zeta=0.43$.
The $B$-line continues for small elongations until the hard sphere limit at $\zeta=0$ is reached. However,
for small elongations $\zeta<\zeta_c = 0.345$, theory predicts a 
novel different scenario: only the translational and the {\it even} rotational 
degrees of freedom freeze at the $B$-line.  Here {\it even} and {\it odd} refer to 
the parity of the order $l$ of the rotational correlator 
$C_{l}(t)=\langle P_{l}(\hat{e}(t)\cdot \hat{e}(0))\rangle$, where $P_{l}$ is the Legendre $l$-polynomial
and $\hat{e}(t)$ is a unity vector along the molecular axis at time $t$.  The {\it odd}-$l$ rotational degrees 
of freedom freeze at higher values of the packing fraction,  namely at an $A$-line $\varphi^{A}_{c}(\zeta)> \varphi_c^B(\zeta)$, 
that merges with the $B$-line at $\zeta=\zeta_c$. The glass transition of the odd degrees 
is of type-$A$, i.e. the long time limit of the odd rotational correlation functions increases 
continuously from zero on crossing the $A$-line.

The $A$- and $B$-lines separate the $\varphi-\zeta$ plane in three dynamic regions: an ergodic fluid, 
a completely arrested state, and an intermediate state (located between the $A$- and $B$-lines) which is 
the amorphous analog of a plastic crystal. In such a state, each molecule remains trapped in the cage
formed by the neighbouring molecules, but is able to perform $180^{\rm o}$ rotations within the cage, 
which lead to relaxation for the odd-$l$ but not for the even-$l$ rotational correlators.
An analogous  scenario is predicted by molecular MCT for the case
of hard-ellipsoids in the limit of small aspect ratio \cite{schilling}.

In order to test these predictions, we have recently carried out molecular 
dynamics simulations \cite{prldumb} in a binary mixture of Lennard-Jones (LJ) dumbbell molecules at fixed packing fraction, 
as a function of the elongation using the temperature $T$ as control parameter (instead of $\varphi$ used in the theoretical work). 
In agreement with MCT predictions, a non-monotonic dependence on the elongation has been observed for the isodiffusivity curves,
with a minimum at a value of $\zeta$ close
to  the maximum of $\varphi_c^B(\zeta)$ \cite{chongtheor1}. 
For small elongations $\zeta \le 0.3$, in deep supercooled states,  
the  coherent intermediate scattering function evaluated at the maximum of the  static structure factor, 
as well as the rotational correlator $C_{2}(t)$, decays to zero at times several orders of magnitude longer than those for $C_{1}(t)$. 
An analysis of the plateau height for the $C_{1}(t)$ correlator shows a sharp drop
of the former in the range $0.3< \zeta <0.5$, consistent with the theoretical $\zeta_c$ value  predicted 
for the hard dumbbell fluid. These features suggest the presence of a 
nearby type-$A$ transition in the investigated LJ dumbbell system.  

Simulations in Ref. \cite{prldumb} have been carried out in
{\it equilibrium} liquid states, for state points above the $B$-transition, i.e., at $T$ such that translational, and both
the odd and even rotational degrees of freedom are ergodic. 
In this manuscript we complement equilibrium simulations and we also investigate the reorientational dynamics 
{\it below} the $B$-transition, i.e., 
in {\it out-of-equilibrium} states, where the translational and even rotational degrees of freedom are arrested. We aim at providing
evidence of the type-$A$ transition, and to shed light on the dynamic features of amorphous analogs to plastic crystals. 
\begin{figure}[tbh]
\vspace{1 mm}
\includegraphics[width=.315\textwidth]{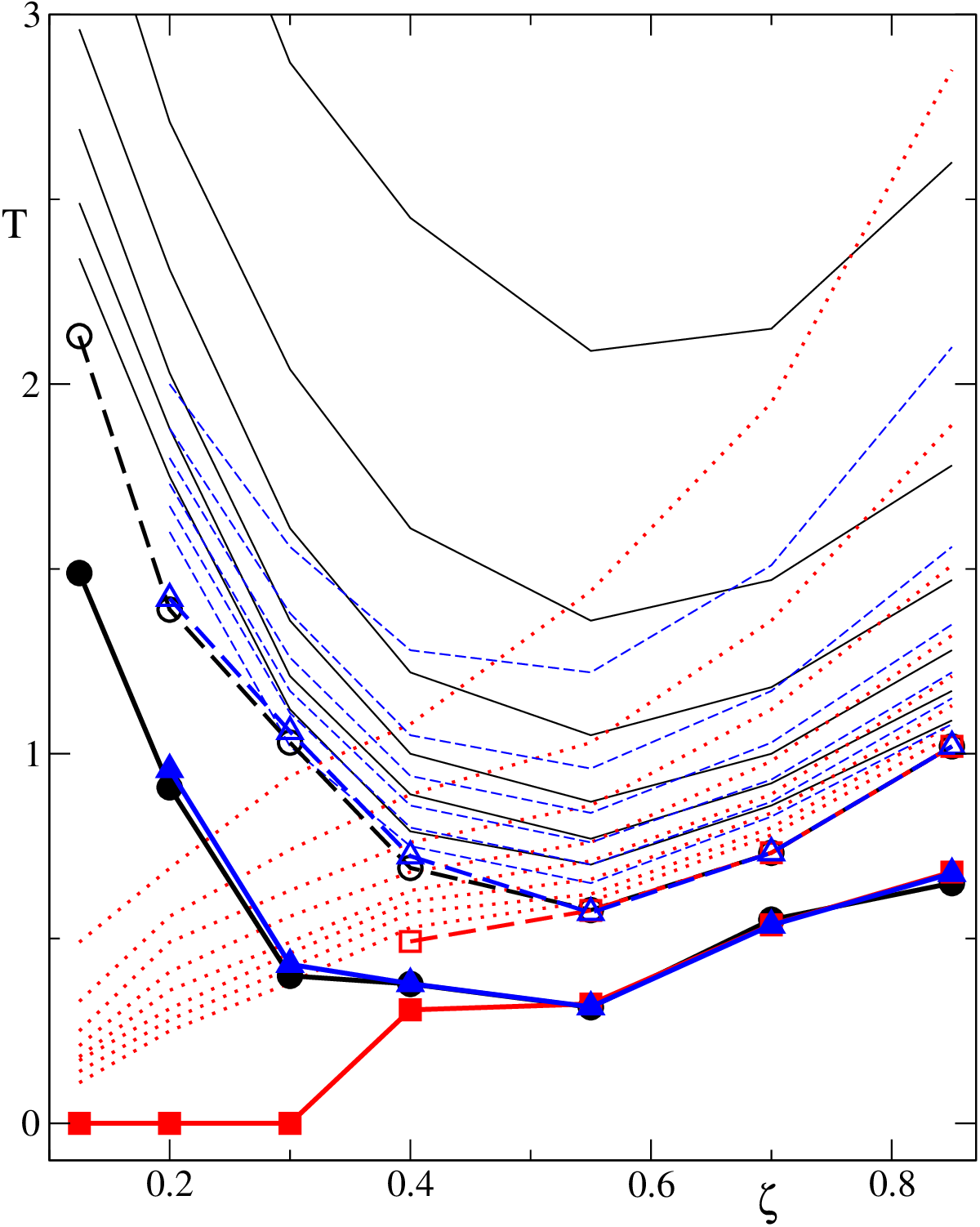}
\vspace{-1 mm}
\caption {Isodiffusivity lines (thin continuous lines), and $\tau_{1}$ (dotted lines) and $\tau_{2}$ (thin dashed lines)
isochrones in the $T$-$\zeta$ plane. These quantities are averaged over all particles in the system.
Line segments join the numerical data (symbols are not included for clarity). The corresponding  values (from top to bottom)
are: $D = 1	\times 10^{-2}, 3 \times 10^{-3}, 1 \times 10^{-3}, 3 \times 10^{-4}, 1	\times 10^{-4}, 3 \times 10^{-5}$; 
$\tau_{1}= 3 \times 10^{1}, 1 \times 10^{2}, 3 \times 10^{2}, 1	\times 10^{3}, 3 \times 10^{3}, 1 \times 10^{4}, 
3 \times 10^{4}, 1 \times 10^{5}$; $\tau_{2}= 3	\times 10^{1}, 1 \times 10^{2}, 3 \times 10^{2}, 1 \times 10^{3},
3 \times 10^{3}, 1 \times 10^{4}$. Empty and filled symbols are respectively the estimated MCT and VFT temperatures (see text)
for $D$ (circles), $\tau_{1}$ (squares), and $\tau_{2}$ (triangles). Thick dashed and thick continuous lines are corresponding guides for the eyes.}
\vspace{-3 mm}
\label{fig:phasediag}
\end{figure}

We investigate a binary mixture of 410:102 dumbbell molecules \cite{notemix} in a box of side $L$ with 
periodic boundary conditions. Each molecule consists of two identical atoms, denoted by A and B 
respectively for the most and the less abundant component of the mixture. The interaction between atoms
of different molecules is given by a Lennard-Jones
potential plus a linear term:
$V_{\alpha\beta}(r) = 4\epsilon_{\alpha\beta}[(\sigma_{\alpha\beta}/r)^{12}
-(\sigma_{\alpha\beta}/r)^{6} +A_{0}+A_{1}(r/\sigma_{\alpha\beta})]$,
where $\alpha$, $\beta \in (A,B)$. The interaction parameters are the same of the monoatomic binary
mixture of Kob and Andersen \cite{koband}, i.e., $\epsilon_{AA}=1.0$,
$\epsilon_{AB}=1.5$, $\epsilon_{BB}=0.5$, $\sigma_{AA}=1.0$,
$\sigma_{AB}=0.8$, $\sigma_{BB}=0.88$. The values of $A_{0}$ and $A_{1}$ are
selected to guarantee continuity of potential and forces at the cutoff distance $c\sigma_{\alpha\beta}$,
with $c=2.5$. The resulting values are $A_{0}=c^{-6}(7-13c^{-6})$ and $A_{1}=6c^{-7}(2c^{-6}-1)$.
The same atomic mass $m=1$ is used for both
A and B particles (each dumbbell has mass equal to two). 

Simulations have been carried out
for elongations $\zeta=$ 0.125, 0.20, 0.30, 0.40, 0.55, 0.70 and 0.85
at different values of $T$ and constant packing fraction $\varphi = 0.708$.
The packing fraction is defined by $\varphi = (\pi/6)(N_{AA}\sigma_{AA}^{3}+N_{BB}\sigma_{BB}^{3})L^{-3}
[1 + (3/2)\zeta -(1/2)\zeta^{3}]$, where $N_{AA}=410$, $N_{BB}=102$ are the number of AA and BB molecules. 
Temperature, time and diffusion coefficient $D$ are respectively
calculated in units of $\epsilon_{AA}/k_{B}$, $(m\sigma_{AA}^{2}/\epsilon_{AA})^{1/2}$ and
$(\sigma_{AA}^{2}\epsilon_{AA}/m)^{1/2}$. Equations of motion are integrated, with a time step 
ranging from  $2\times 10^{-4}$ to $5 \times 10^{-3}$ according to $T$ and/or $\zeta$, using the velocity Verlet scheme 
combined with the SHAKE algorithm \cite{allen} for keeping the molecular elongation constant. 
At each state point, the system is thermalized at the requested $T$ by periodic velocity rescaling. 
A microcanonical run is performed after suitable equilibration to generate a representative trajectory in phase space.  
Statistical averages are performed over 20 independently generated equilibrium trajectories. At high $T$, the 
thermalization process is started from a distorted crystalline configuration
that quickly melts. At low $T$, it is started from equilibrated high $T$ configurations.
After reaching equilibrium, energy or pressure show no drift. Also, translational and rotational 
mean squared displacements as well as dynamic correlation functions,
show no aging, i.e. there is no systematic dependence of the
dynamical properties from the selected time origin.
In practice equilibration requires that 
each dumbbell, in average, has moved several diameters and performed several full rotations from its initial position. 

At very low $T$ and small $\zeta$, relaxation times for translational an even rotational degrees of 
freedom become exceedingly long for the time scale of the simulation.
In these conditions only the odd rotational degrees of freedom decay to zero. We perform long 
runs of $\approx 10^7$ steps until neither significant drift in $T$ and pressure, nor aging in the 
translational and even rotational correlation functions is observed, to ensure that the residual 
rotational motion of the dumbbells takes place in an essentially frozen structure \cite{aging}.  
After this preliminary simulation, trajectories are calculated. 
\begin{figure}[tbh]
\includegraphics[width=.315\textwidth]{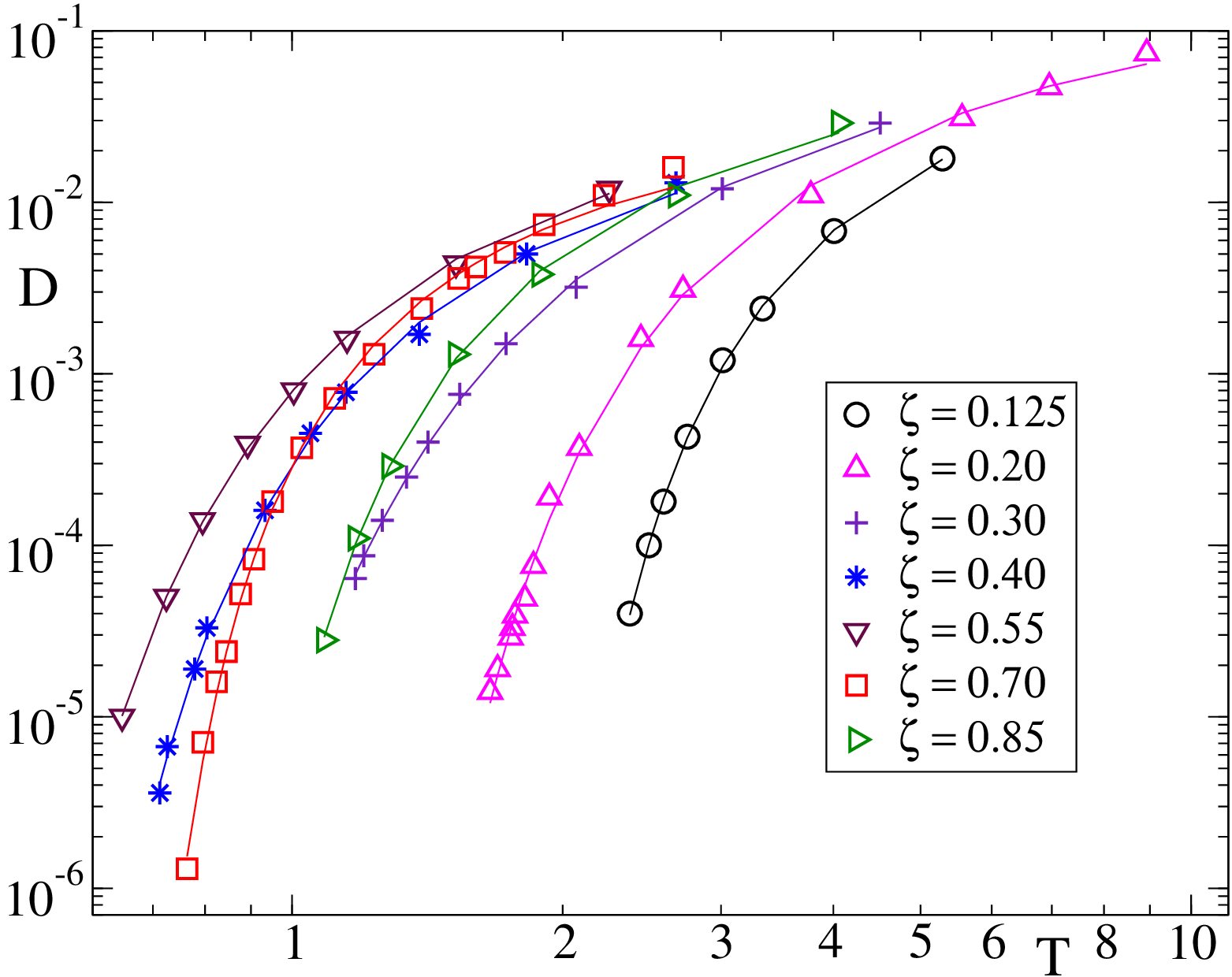}
\newline
\newline
\includegraphics[width=.315\textwidth]{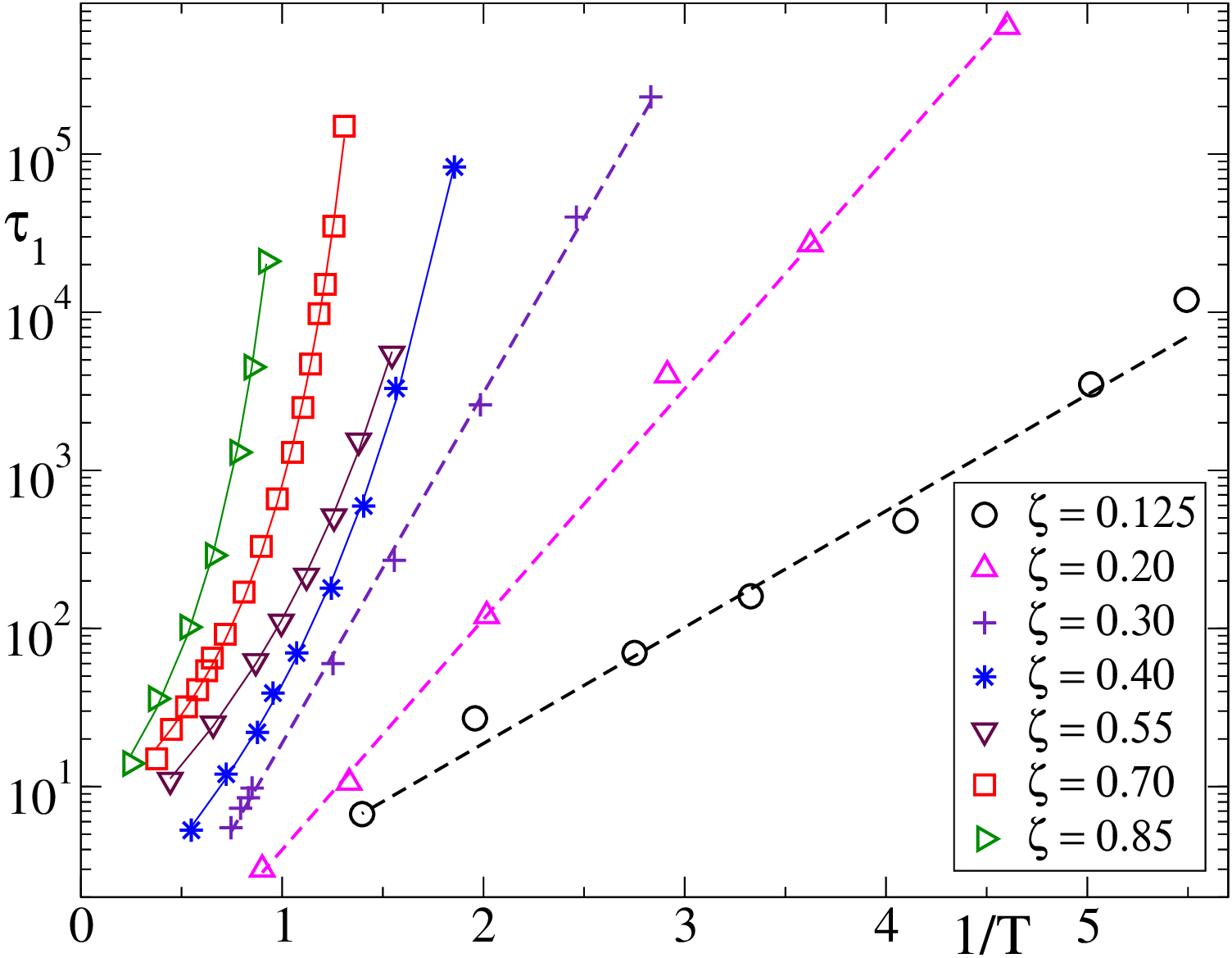}
\newline
\newline
\includegraphics[width=.315\textwidth]{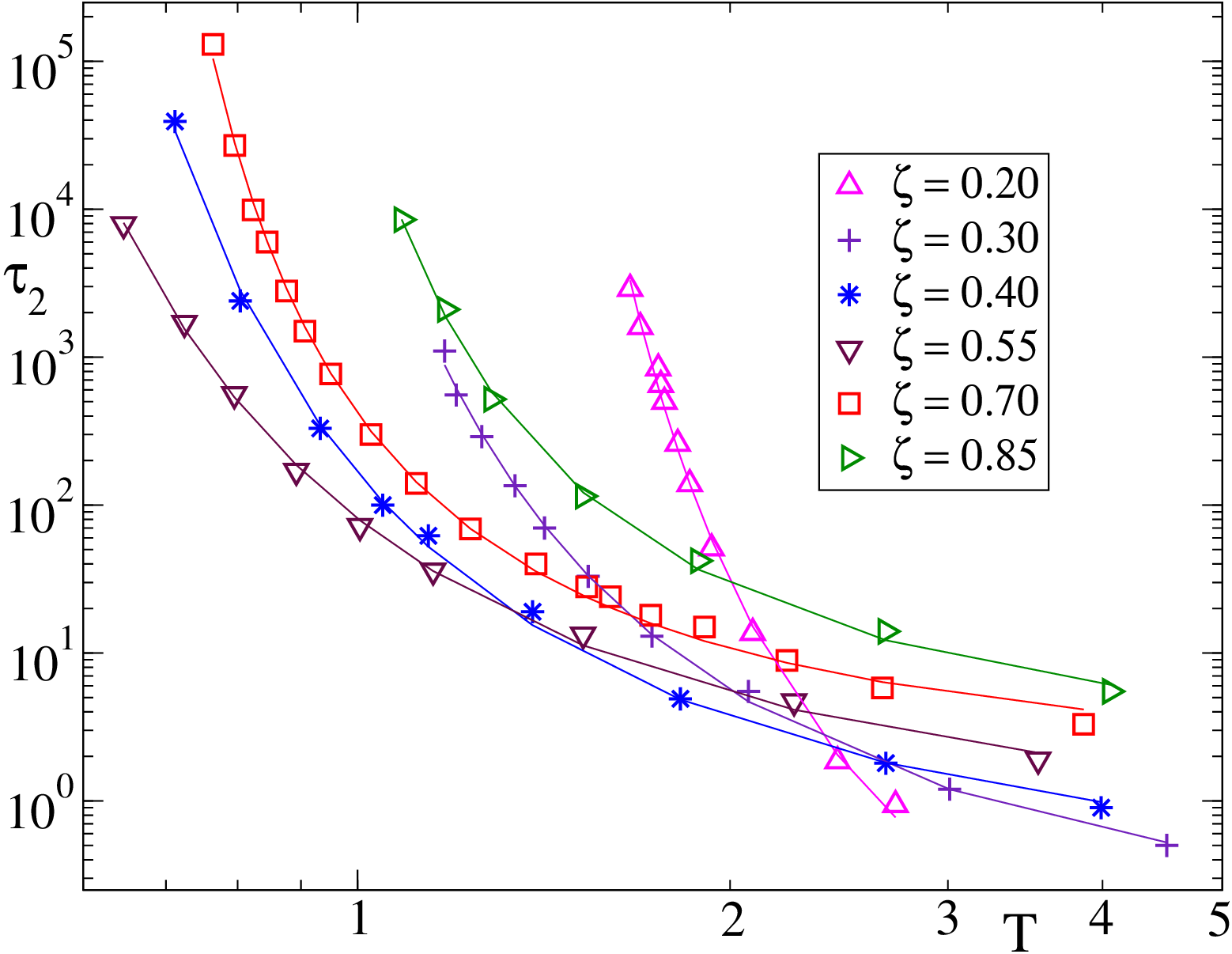}
\newline
\vspace{-3 mm}
\caption{Symbols: Temperature dependence  of $D$, $\tau_{1}$ and $\tau_{2}$.
Continuous lines are fits to VFT functions (see text). Dashed lines are Arrhenius fits.
Data for $\tau_{1}$ have been represented vs. $1/T$ in order to evidence the small $\zeta$  Arrhenius
behavior.}
\vspace{-4 mm}
\label{fig:vft}
\end{figure}

Fig. \ref{fig:phasediag} shows data for  $D$ and for  the relaxation times $\tau_{1}$, $\tau_{2}$ of the 
rotational correlators $C_{1}(t)$ and $C_{2}(t)$.
The diffusion coefficient is evaluated from the long time
limit of the mean square displacement,  while
$\tau_{1}$ and $\tau_{2}$ are defined as the time at which the corresponding correlation 
functions decay to  3 \% of their initial value.  
As previously done in the analysis of other systems  characterized by slow dynamics \cite{arrest,prldumb}, 
we report isochrones (constant $\tau_{l}$)  and isodiffusivity
(constant $D$) curves, to provide an indication of the shape of the glass line in the $(T-\zeta)$ plane. 
The isodiffusivity curves are consistent with the same quantities calculated 
in Ref. \cite{prldumb} and, as previously noted,
show a mimimum around $\zeta= 0.55$, close to the MCT prediction for hard dumbbells \cite{chongtheor1}. 
Fig. \ref{fig:phasediag} shows that, especially at low $T$, the iso-$\tau_{2}$
curves follow the same  trend as the isodiffusivity curves, tending to collapse
in a common curve, as expected 
on approaching a type-$B$ transition. 
The iso-$\tau_{1}$ curves  show a similar trend only at large $\zeta$. On the contrary, for small elongations 
the iso-$\tau_{1}$ curves depart from the iso-$D$ and iso-$\tau_{2}$ curves.  Differently from $\tau_2$, 
within the resolution in $\zeta$ of our calculations, the $\tau_1$ isochrones are monotonic in the $T-\zeta$ plane.  
The fact that the iso-$\tau_1$ curves lay significantly below the
iso-$\tau_2$ curves shows very clearly that fast relaxation of the $odd$ degrees of freedom takes 
place even when the even degrees are arrested.

A detailed quantitative comparison of the presented results with MCT predictions for dynamic exponents 
and scaling laws can not be simply performed, except for large $\zeta$. 
Indeed, it has been shown that the range of validity of the asymptotic
MCT laws --- for the small $\zeta$ hard dumbbell fluid case \cite{chongtheor2} ---
is limited to unusually small values of the separation parameter $\epsilon=(\varphi_{c}^{B}-\varphi)/\varphi_{c}^{B}$. 
In particular, the exponent $\gamma$ for the well-known power law
$D, 1/\tau \propto (\varphi_{c}^{B}-\varphi)^{\gamma}$ can only be correctly 
estimated by exploring the  range $\epsilon < 10^{-2}$. For small $\zeta$, a fit of data 
in the range $10^{-2} < \epsilon < 10^{-1}$ would show an  apparent power law behavior, but the corresponding fit
would yield an ``effective'' exponent, which contrary to the asymptotic limit, would  not take a
universal value, but would depend on the chosen observable \cite{chongtheor2}. 
A similar phenomenon is expected to be present in 
the studied system at small $\zeta$, preventing the possibility of
an accurate determination of the asymptotic laws close to the MCT critical temperature $T_{c}^{B}$. In addition, we recall 
that ideal MCT neglects activated processes which are known to be present in reality around and below $T_c^{B}$ \cite{hopping}.  
Only when hopping  phenomena can be neglected, an MCT study can be feasible very close to $T_c^{B}$.
%

Fig. \ref{fig:vft} shows the $T$ dependence
of $D$, $\tau_1$ and $\tau_2$ for several $\zeta$ values.
If a fit (not shown) of the $T$ dependence of $D$ and $1/\tau_{2}$  according to
the functional form $(T-T_c^{B})^{\gamma}$ is performed, we obtain very similar values of $T_c^{B}$ for both quantities 
in all the $\zeta$ range
(see Fig. \ref{fig:phasediag}). For the case of $1/\tau_1$, reasonable power-law fits
can be performed only for $\zeta \gtrsim 0.4$.  For  $ \zeta = 0.55  $ fits provide 
values of $T_c^{B}$ consistent with those obtained for $D$ and $1/\tau_2$  but with different $\gamma$ values. 
Only for $\zeta \ge 0.70$ a simultaneous consistent description 
for $D$, $1/\tau_{1}$ and $1/\tau_{2}$ is achieved, with identical $T_c^{B}$ and $\gamma$ values. 
Specifically  $T_{c}^{B}=0.732$, $\gamma=2.50$ for $\zeta=0.70$ 
and  $T_{c}^{B}=1.02$, $\gamma=2.06$ for $\zeta=0.85$.

To provide a lower bound to $T_c^{B}$ at small $\zeta$, we also fit the data in terms of the well-known phenomenological 
Vogel-Fulcher-Tammann (VFT) law $D, 1/\tau_{1,2} \propto \exp[-A/(T-T_{VFT})]$, which predicts a dynamic arrest 
of the corresponding degree of freedom at a finite temperature $ T_{VFT} < T_{c}^{B}$ \cite{hopping}. In all cases, the VFT law
provides a consistent effective representation of the dynamics, with a fit quality comparable to the
MCT power law fit. The resulting $T_{VFT}(\zeta)$ for $D$, $\tau_1$ and $\tau_2$ are 
reported in Fig. \ref{fig:phasediag}. Interestingly,
$T_{VFT}$ for $D$, $\tau_{1}$ and $\tau_{2}$  coincide
for large elongations, indicating strong couplings between the 
translational and all the rotational degrees of freedom. However, for $\zeta \lesssim 0.4$
the $T_{VFT}$ for $\tau_{1}$
is much smaller than  the $T_{VFT}$ for $D$ and $\tau_{2}$,
confirming that around this value of $\zeta$ the odd degrees of freedom completely decouple form the even and 
from the translational ones. This latter value of $\zeta$ is consistent  with the merging point 
of the $A$- and $B$-transitions for the hard dumbbell fluid \cite{chongtheor1}
and with estimates of Ref. \cite{prldumb}.
At smaller elongations $\zeta  \le 0.3$, for $\tau_{1}$ we find  
$T_{VFT} \approx 0$. This implies that 
a good description of the data is provided by an Arrhenius
law $\tau_{1} \propto \exp(E/k_{B}T)$ (see Fig. \ref{fig:vft}).
The activation energies $E/\epsilon_{AA}=$ 1.8, 3.3 and 5.0, respectively for $\zeta$= 0.125, 0.20 and 0.30,
roughly follow a linear dependence on $\zeta$.
\begin{figure}
\hspace{7mm}
\includegraphics[width=.305\textwidth]{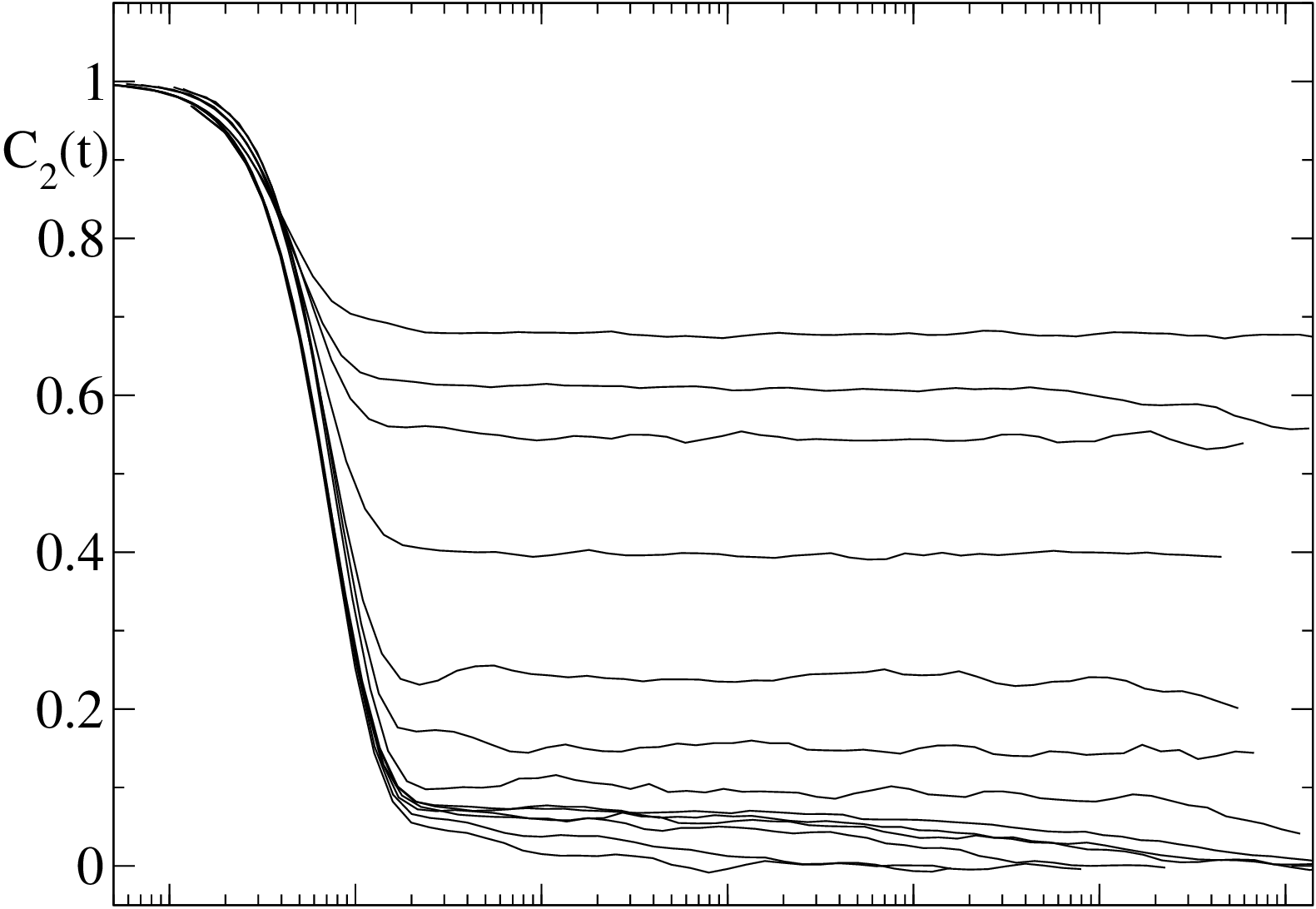}
\newline
\newline
\vspace{-2.5 mm}
\includegraphics[width=.305\textwidth]{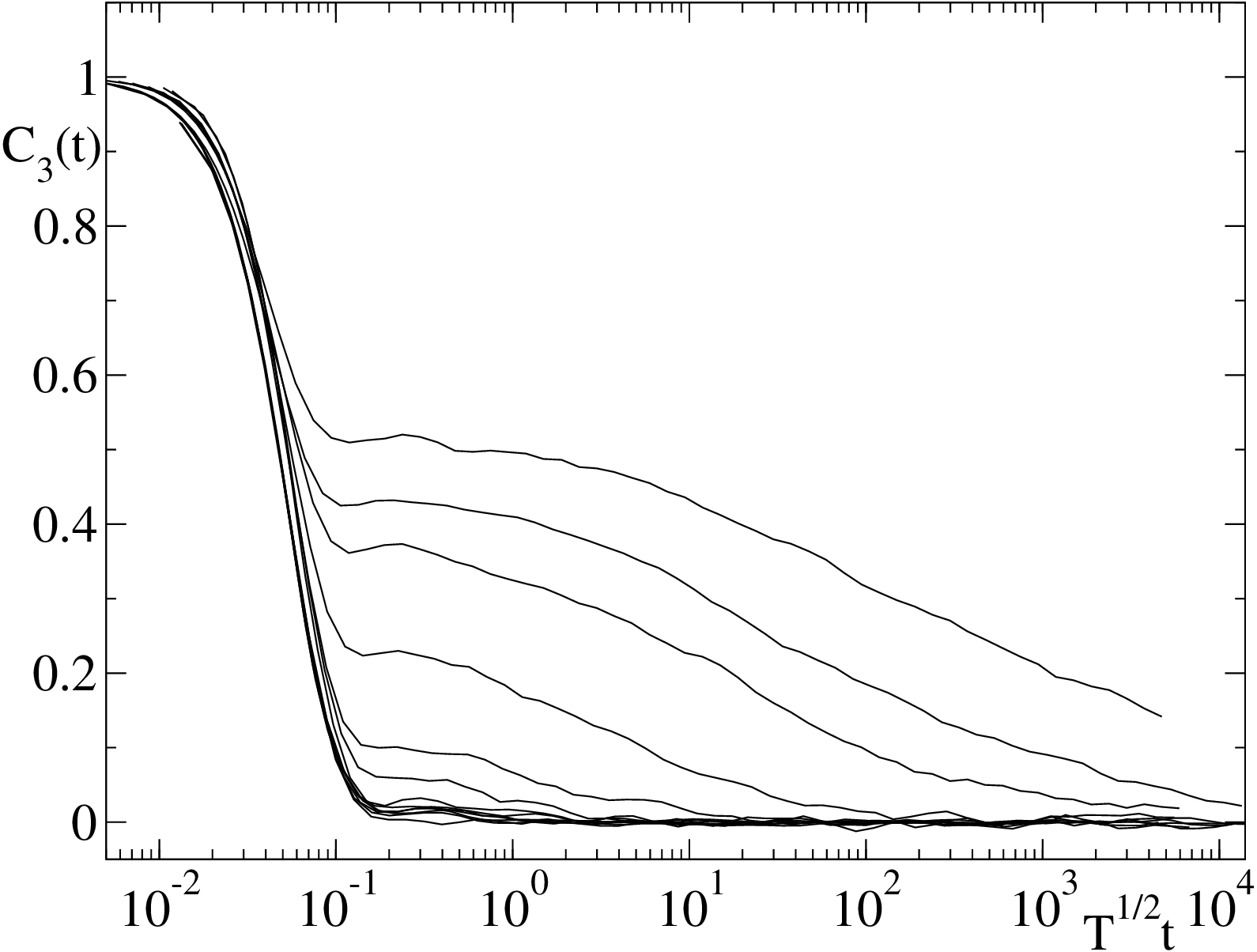}
\vspace{-1 mm}
\caption{Temperature dependence of the rotational correlators $C_{2}(t)$ and $C_{3}(t)$ vs. time
rescaled by $T^{1/2}$, for elongation $\zeta=0.2$. 
From top to bottom $T$= 0.217, 0.276, 0.343, 0.496, 0.750, 1.110, 1.396,
1.693, 1.770, 1.815, 1.933, 2.443 and 3.776.}
\vspace{-3 mm}
\label{fig:c2c3}
\end{figure}
The presence of an activated Arrhenius $T$ dependence in
$\tau_1(T)$   for small molecular elongations, strongly suggests that the residual reorientational 
dynamics at very low temperatures, in the $T$-region where the translational dynamics is already arrested, 
is characterized by $180^{\rm o}$ rotations which
do not lead to relaxation of the even rotational degrees of freedom. The Arrhenius functional form also suggests that
hopping phenomena are extremely relevant, so much to
completely mask the presence of a finite type-$A$ glass transition
temperature, if the analysis is limited only to the $T$ dependence of $\tau_1$ (or any other odd $\tau$).
%

While we can not expect, due to this intense hopping decorrelation, to observe any critical law, 
indication of a residual cross-over finite temperature can still be
observed in the full dependence of the correlation functions, and in particular  in the strength of 
the  long time $\alpha$ relaxation process.  The main effect of hopping is
indeed manifested in the long time decay from the plateau value. Indeed,  in the absence of these 
additional relaxation channels, the correlation function  would decay only down to the plateau value.  
Fig. \ref{fig:c2c3} shows,  for $\zeta=0.2$,  the $T$-dependence of $C_{2}(t)$ and $C_{3}(t)$, 
in a wide range of $T$,  (across $T_{VTF} = 0.96 $ for $\tau_{2}$). Though $C_{1}(t)$ has more 
experimental relevance, $C_{3}(t)$ is more adequate to discuss the behavior of the non-ergodicity parameter, 
since it shows a more pronounced  plateau. A representation  of $C_{i}(t)$ vs. $tT^{1/2}$, i.e rescaling the time  
by the thermal velocity, yields a  clearer visualization of the decay of the correlation function. Data reported in 
Fig. \ref{fig:c2c3} show two important  features:

(i) at low $T$ the plateau height (i.e., the non-ergodicity factor) for both  $C_2$ and $C_3$ progressively 
grows up, signaling a crossing of the critical MCT temperatures.
However, while in case of $C_{2}(t)$, the increase of the plateau value is connected to a non-ergodic long 
time behavior (i.e. $C_2$ does not decay to zero within the simulation time window), $C_{3}(t)$ always  relaxes toward zero.
This fact evidences that hopping processes restore ergodicity for reorientations at $T$ far below the 
temperature at which the plateau value starts to increase from zero.
(ii) The decay of the  $C_2$ correlation function is characterized by a finite $T$ interval in which the plateau 
value is essentially constant (within the present numerical error). For smaller $T$, the height of the 
plateau starts to increase, as discussed above. Instead, the decay of the $C_3$ correlation function is not 
characterized by any plateau at high $T$, and the increase of the plateau starts from zero, consistently 
with the presence of a type-$A$ transition which, due to the hopping effects, is significantly weakened. 
Moreover, the first $T$ at which the $C_3$ plateau value is different from zero is lower than the $T$ at 
which the $C_2$  plateau starts to increase. This suggests that the type-$A$ 
transition temperature 
is different and lower than the type-$B$ $T_c^{B}$. We also note that the
$\alpha$-relaxation process for odd $C_i$ is highly stretched, which might reflect the
local cage heterogeneity.

In summary, we have provided evidence for the existence of the type-A transition predicted
by MCT for the reorientational dynamics of glass-forming liquids of small molecular elongation.
Our numerical results complement those recently reported for $T>T_c^B$ in Ref. \cite{prldumb}.
By computing relaxation times and plateau heights of odd rotational dynamic correlators
--- in the arrested state for the even rotational and the translational degrees of freedom--- we show that 
reorientational hopping dominates the dynamics and restores ergodicity
down to temperatures far below the type-$A$ transition. Though these results have been 
specifically obtained for a liquid of symmetric dumbbells, experimental techniques
probing  $C_{1}(t)$ and $C_{2}(t)$, at very low temperatures, will yield qualitatively similar results
for small elongated molecules with a weak asymmetry
which disfavours rotations different from $\approx 180^{\rm o}$. 
Finally, the present model can become a test case for developing theories of the glass transition in which
activated processes --- which become relevant around and below $T_c^{B}$ --- can be studied.
\\  
MIUR-FIRB is acknowledged for financial support.

%

%
%
%

\begin{thebibliography}{99}
\bibitem{notemix}
The system investigated in Ref. \cite{prldumb} is a 800:200
mixture with a modified LJ interaction.
\bibitem{aging}
In principle, signifficant aging effects for translational and even rotational degrees of freedom
could be observed for much longer simulation times,
beyond the limit of current computational resources.
\end{thebibliography}
\end{document}